\documentclass[a4paper]{jpconf}

\usepackage{graphicx}
\usepackage{bm}
\usepackage{color}
\usepackage{braket}
\usepackage{amsfonts}
\usepackage{amsmath}

\def\dblone{\hbox{$1\hskip -1.2pt\vrule depth 0pt height 1.6ex width 0.7pt
     \vrule depth 0pt height 0.3pt width 0.12em$}}

\begin{document}
\title{Microscopic Optical Potentials: recent achievements and future perspectives}

\author{Paolo Finelli}

\address{Dipartimento di Fisica ed Astronomia, Universit\`a degli Studi di Bologna and INFN, Sezione di Bologna, Via Irnerio 46, I-40126 Bologna, Italy}

\ead{paolo.finelli@unibo.it}

\author{Matteo Vorabbi}

\address{Department of Physics, University of Surrey, Guildford, Surrey, GU2 7XH, UK}

\ead{m.vorabbi@surrey.ac.uk}

\author{Carlotta Giusti}

\address{Dipartimento di Fisica, Universit\`a degli Studi di Pavia and INFN, Sezione di Pavia, I-27100 Pavia, Italy}

\ead{carlotta.giusti@pv.infn.it}

\begin{abstract}
Few years ago we started the investigation of microscopic Optical Potentials (OP) in the framework of chiral effective field theories \cite{Epelbaum_review, Machleidt_review} and published our results in a series of manuscripts. Starting from the very first work \cite{Vorabbi2016}, where a microscopic OP was introduced following the multiple scattering procedure of Watson \cite{Watson}, and then followed by Refs. \cite{VorabbiN4LO, VorabbiTalys}, where the agreement with experimental data and phenomenological approaches was successfully tested, we finally arrived at a description of elastic scattering processes off non-zero spin nuclei \cite{VorabbiSpin}.  Among our achievements, it is worth mentioning the partial inclusion of three-nucleon forces \cite{Vorabbi3B}, and the extension of our OP to antiproton-nucleus elastic scattering \cite{VorabbiAnti}. Despite the overall good agreement with empirical data obtained so far, we do believe that several improvements and upgrades of the present approach are still to be achieved.

In this short essay we would like to address some of the most relevant achievements and discuss an interesting development that, in our opinion, is needed to further improve microscopic OPs in order to reach in a near future the same level of accuracy of the phenomenological ones. 
\end{abstract}

\section{Introduction}
It is well known that a suitable and successful framework to describe nucleon-nucleus processes is provided by the concept of the nuclear optical potential. 
Within this approach it is possible to compute the scattering observables across wide regions of the nuclear landscape and to extend calculations to inelastic channels or to a wide variety of nuclear reactions, e.g., nucleon transfer, knockout, capture, or breakup.
Even if it is true that a phenomenological approach is generally preferred to achieve a more accurate description of the available experimental data, nowadays, with the upcoming facilities for exotic nuclei (FRIB at MSU just to mention one of the most important projects \cite{FRIBpaper}), we strongly believe that a microscopic approach will be the preferred way to make robust predictions and to assess the unavoidable theoretical uncertainties already in a near future \cite{FRIB_workshop}.

\section{Rooting optical potentials within \textit{ab initio} approaches}

The calculation of a microscopic optical potential requires, in principle, the solution of the full many-body nuclear problem for the incident nucleon and the A nucleons of the target. In practice, with suitable approximations, microscopic optical potentials are usually calculated from two basic quantities: the
nucleon-nucleon ($NN$) $t$ matrix and the matter distribution of the nucleus in the coordinate $\rho({\bm r},{\bm r^\prime})$, or in the momentum $\rho({\bm k},{\bm k^\prime})$ representation space. 
Formally we have to start from the 
the full $(A+1)$-body Lippmann-Schwinger equation that allows to determine the two-body transition matrix $T$ defined as
\begin{equation}
V | \Psi \rangle = T | \phi \rangle \; ,
\end{equation}
where $|\Psi \rangle$ is the scattered state and $|\Phi \rangle$ is a free wave (asymptotic) solution,
in terms of an external two-body interaction $V$
\begin{equation}
\label{generalscatteq}
T = V + V G_0 (E) T \, .
\end{equation}
The previous relation can be manipulated in order to obtain a set of two equations in terms of an auxiliary quantity, i.e. the optical potential $U$. 

The first approximation we need is given 
by the spectator expansion \cite{Weppner1995}, in which the scattering relation is expanded in a finite series of terms where the target nucleons interact directly with the incident proton. In particular, the first term of this series only involves the interaction of the projectile with a single target nucleon, the second term involves the interaction of the projectile with two target nucleons, and so on to the subsequent orders.

As a consequence we can split Eq. (\ref{generalscatteq}) into a set of two equations.
The first one is an integral equation for $T$
\begin{equation}\label{firsttamp}
T = U + U G_0 (E) P T \, ,
\end{equation}
where $U$ is the optical potential operator, and the second one is an integral equation for $U$
\begin{equation}\label{optpoteq}
U = V + V G_0 (E) Q U \, .
\end{equation}
$G_0 (E)$ is the free propagator for the $(A+1)$-nucleon system and 
the operators $P$ and $Q$ in Eqs.~(\ref{firsttamp}) and (\ref{optpoteq}) are projection operators,
\begin{equation}
P + Q = \dblone \, .
\end{equation}
For elastic scattering processes $P$ selects only the ground state $|\Phi_0\rangle$, i.e.
\begin{equation}
P = |\Phi_0 \rangle \langle \Phi_0| \; .
\end{equation}
Eq. (\ref{optpoteq}) can be numerically solved if we assume the validity of another approximation, namely the impulse approximation where, since the projectile energy scales involved are large compared to nuclear binding forces of the interacting target nucleon, we can safely use the free $NN$ $t$ matrix in the Eq. (\ref{firsttamp}). Most of the theoretical derivations can be found in Refs. \cite{Vorabbi2016,Weppner1995}.


\begin{figure}
\begin{center}
\includegraphics[scale=0.60]{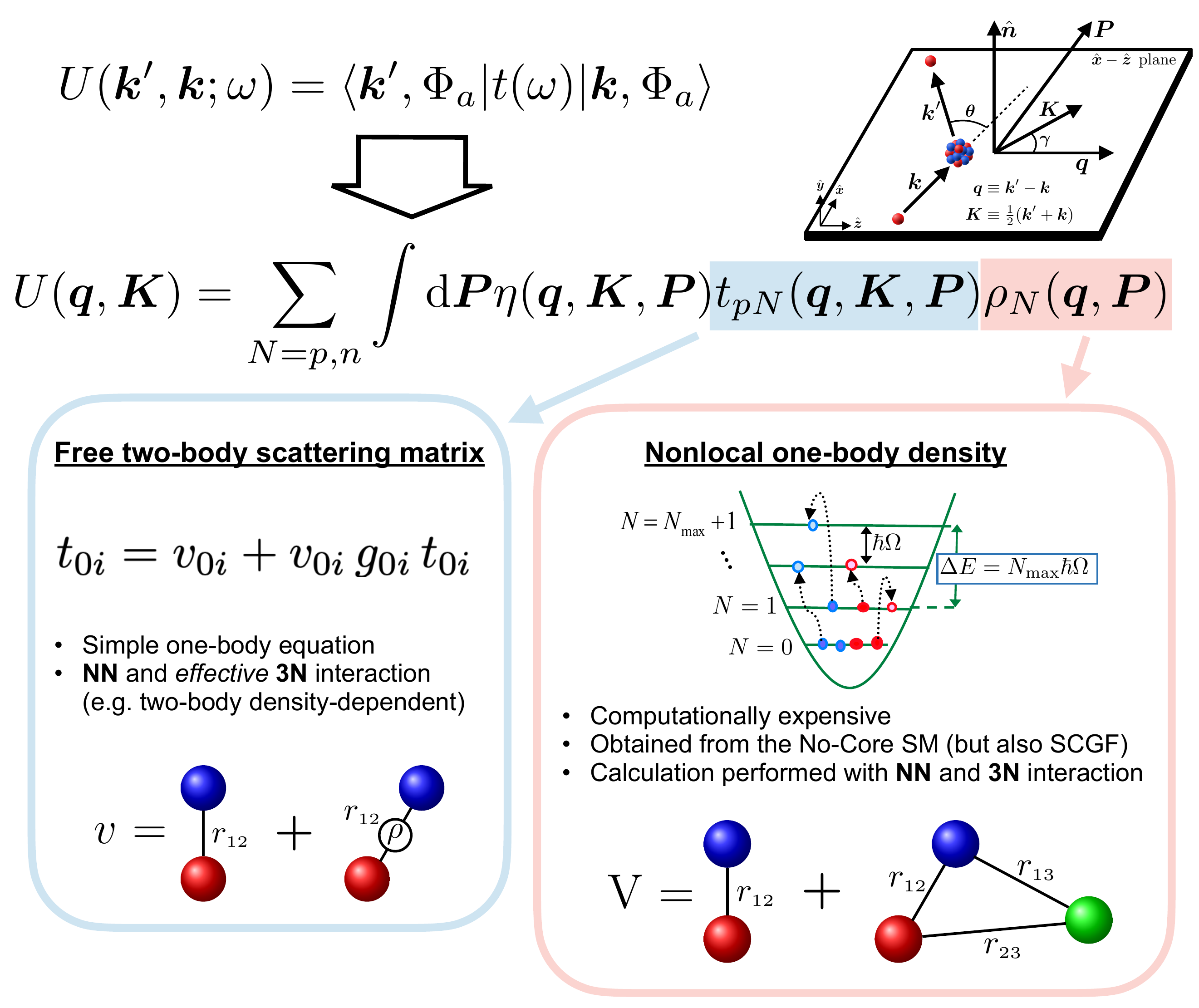}
\end{center}
\caption{\label{fig1}
In this {\it infographic} the main building blocks of the optical potential are schematically shown along the main features of each components. For practical reasons the auxiliary optical potential $U$ is expressed as a function of the center-of-mass momentum ${\bm K} \equiv 1/2({\bm k} + {\bm k}^\prime)$, and the momentum transfer ${\bm q} \equiv ({\bm k} - {\bm k}^\prime)$.
It can be factorized into three pieces \cite{PhysRevC.40.881,PhysRevC.41.814, PhysRevC.56.2080, PhysRevC.57.1378}: the M\"oller correction $\eta({\bm q}, {\bm K}, {\bm P})$ that ensures the conservation of the transmitted flux after the transformation in the center-of-mass reference frame, the {\it projectile-target} potential $t_{NN}({\bm q}, {\bm K};\omega)$ that depends also on the energy of the projectile $\omega$, and the {\it structure} component $\rho_N({\bm q}, {\bm P})$, i.e. the Fourier transform of the non-local matter density in the momentum space. For both inputs we show into the coloured insertions the main features. For the potential part, at the moment, we are able to include the full $NN$ interaction and averaged $NNN$ contributions in terms of a medium density-dependent $NN$ interaction \cite{Holt}. 
This solution requires to fix a density $\bar{\rho}$ at which the calculations are performed. So far \cite{VorabbiSpin,Vorabbi3B} we decided to explore the impact of $NNN$ terms performing calculations for several values of $\bar{\rho}$ (usually in the range $0.08$ fm$^{-3} \le \bar{\rho} \le 0.13$ fm$^{-3}$, see Figs 1-6 of Ref. \cite{VorabbiSpin}). Investigations in this direction are currently underway \cite{3bpaper}.
For the structure part we used the NCSM \cite{BARRETT} for light nuclei that allows us to fully use both $NN$ and $NNN$ terms (see Sect II and III. of Ref. \cite{VorabbiSpin}). For medium nuclei other approaches should be explored, like Self-Consistent Green Functions (SCGF) \cite{Barbieri_review1,Barbieri_review2}. 
}
\end{figure}


Then, from a practical point of view, the OP can be computed in momentum space as follows (after some mathematical manipulations that can be found in Refs.~\cite{PhysRevC.40.881,PhysRevC.41.814, PhysRevC.56.2080, PhysRevC.57.1378})
\begin{equation}\label{fullfoldingop}
\begin{split}
U ({\bm q},{\bm K}; E ) = \sum_{N=p,n} &\int d {\bm P} \; \eta ({\bm q},{\bm K},{\bm P}) \,
t_{NN} \left[ {\bm q} , \frac{1}{2} \left( \frac{A+1}{A} {\bm K} + \sqrt{\frac{A-1}{A}} {\bm P} \right) ; E \right] \\
&\times \rho_N \left( {\bm P} + \sqrt{\frac{A-1}{A}} \frac{{\bm q}}{2} , {\bm P} - \sqrt{\frac{A-1}{A}} \frac{{\bm q}}{2} \right) \, ,
\end{split}
\end{equation}
where ${\bm q}$ and ${\bm K}$ represent the momentum transfer and the average momentum, respectively. Here ${\bm P}$ is an integration variable, $t_{NN}$ is the $NN$
$t$ matrix and $\rho_N$ is the one-body nuclear density matrix. 
The parameter $\eta$ is the M\"{o}ller factor, that imposes the Lorentz invariance of the flux when we pass from
the $NA$ to the $NN$ frame in which the $t$ matrices are evaluated. 
Finally, $E$ is the energy at which the $t$ matrices are evaluated and it is fixed at one half 
the kinetic energy of the incident nucleon in the laboratory frame. 

As shown in Fig. \ref{fig1}, the calculation of Eq. (\ref{fullfoldingop}) requires two external inputs: 
the $NN$ $t$ matrix and the one-body density $\rho_N$.

The calculation of the density matrix could be performed using different \textit{ab initio} approaches. The method followed in Refs.~\cite{Burrows,Gennari:2017yez}, where one-body translationally invariant densities were computed within the
{\it ab initio} No-Core Shell Model (NCSM)~\cite{BARRETT} approach proved to be very successful in our own works \cite{VorabbiSpin,VorabbiAnti}. Alternatively other methods could be used, i.e. Self-Consistent Green Functions (SCGF) \cite{Barbieri_review1, Barbieri_review2}, that allow to overcome the main limitation of the NCSM method, namely the severe limits on the atomic mass of the target nucleus due to the computational overload. Work in this direction is currently underway and a thorough investigation of nickel and calcium isotopes will soon appear \cite{Iso_paper}. 
For heavier nuclei, the only available method is Density Functional Theory (DFT). Even if most of the parametrizations have a phenomenological derivation, in recent years some authors \cite{Marino:2021xyd} started to employ such approach from an {\it ab initio} point of view. 
  
The main advantages of using {\it ab initio} methods in our framework is twofold. On one hand we can preserve the consistency of our theoretical framework if the same realistic interactions are used both for the {\it structure} part and the {\it projectile-target} interaction (with some caveats explained in Refs. \cite{Vorabbi2016,VorabbiN4LO, VorabbiTalys, Vorabbi3B, VorabbiSpin}) and, on the other hand, we are able to provide more theoretically founded predictions for exotic nuclei since we do not have to rely on any fitting procedure on any selections of stable nuclei.

To calculate the {\it projectile-target} potential we rely on modern $NN$ potentials derived in the framework of Chiral Perturbation Theory (ChPT). Among the many available choices of realistic potentials, i.e. those that reproduce phase-shifts and binding energies of light nuclei with $\chi^2|_{\rm data} \simeq 1$, the choice of chiral potentials over phenomenological ones responds to some important requirements that will be listed shortly below.

Generally speaking, ChPT is a perturbative technique for the description of hadron scattering amplitudes
based on expansions in powers of a parameter that can be generally defined as $(p,m_\pi)/\Lambda_b$, where $p$ is the magnitude of three-momenta of the external particles, $m_\pi$ is the pion mass, and $\Lambda_b$ is the chiral symmetry breaking scale  \cite{Scherer:2002tk}. 
ChPT respects the low-energy 
symmetries of Quantum ChromoDynamics (QCD) and, up to a certain extent, is model independent and 
systematically improvable by an order-by-order expansion, 
with controlled uncertainties from neglected higher-order terms \cite{Georgi:1994qn}.
Both the fact that ChPT is directly connected to QCD and that the $NN$ components (and also $NNN$) are organized in terms of a systematic theoretical expansion is a huge improvement respect to conventional $NN$ potentials, like CD-Bonn \cite{Machleidt:2000ge} or Nijmegen \cite{nijmegen}. The hierarchy of terms could also be exploited in order to provide an estimate for the theoretical uncertainties. 

From a practical point of view, for our own calculations \cite{Vorabbi2016,VorabbiN4LO,VorabbiTalys,VorabbiSpin,Vorabbi3B,VorabbiAnti} we mainly employed the $NN$ chiral interactions developed by Entem {\it et al.}~\cite{Entem:2014msa,Entem:2017gor} up to the fifth order (N$^4$LO) or, alternatively, the ones developed by Epelbaum {\it et al.} \cite{Epelbaum:2014sza,Epelbaum:2014efa} following a different strategy. The main difference between the two theoretical approaches of the chiral potentials concerns the renormalization procedures.
The strategy followed in Refs. \cite{Epelbaum:2014sza,Epelbaum:2014efa} consists in a coordinate space regularization 
for the long-range contributions, and a conventional momentum space regularization for the contact (short-range) terms. 
On the other hand, for Refs. \cite{Entem:2014msa, Entem:2017gor}, a slightly more conventional approach was pursued. 
A spectral function regularization was employed to regularize the loop contributions and a conventional regulator function to deal with divergences in the Lippman-Schwinger equation. 

The three-nucleon forces deserve a separate treatment since so far we can employ genuine $3N$ forces only to compute the one-body densities of the target nuclei (up to the third order N\textsuperscript{2}LO with a local and nonlocal regularization~\cite{Navratil2007,Gysbers2019}). 
At N\textsuperscript{2}LO order we include the $2\pi$ exchange diagram between three nucleons, a one-$\pi$-exchange plus a $NN$ contact term and a $3N$ contact term. 
For more details and an explicit derivation of the relevant formulae, we refer the reader to Refs. \cite{ Navratil2007}.
For the {\it projectile-target} part we follow the prescriptions by Holt {\it et al.} \cite{Holt} to derive a suitable density-dependent NN interaction.

\subsection{A selection of results}

\begin{figure}
\begin{center}
\includegraphics[scale=0.5]{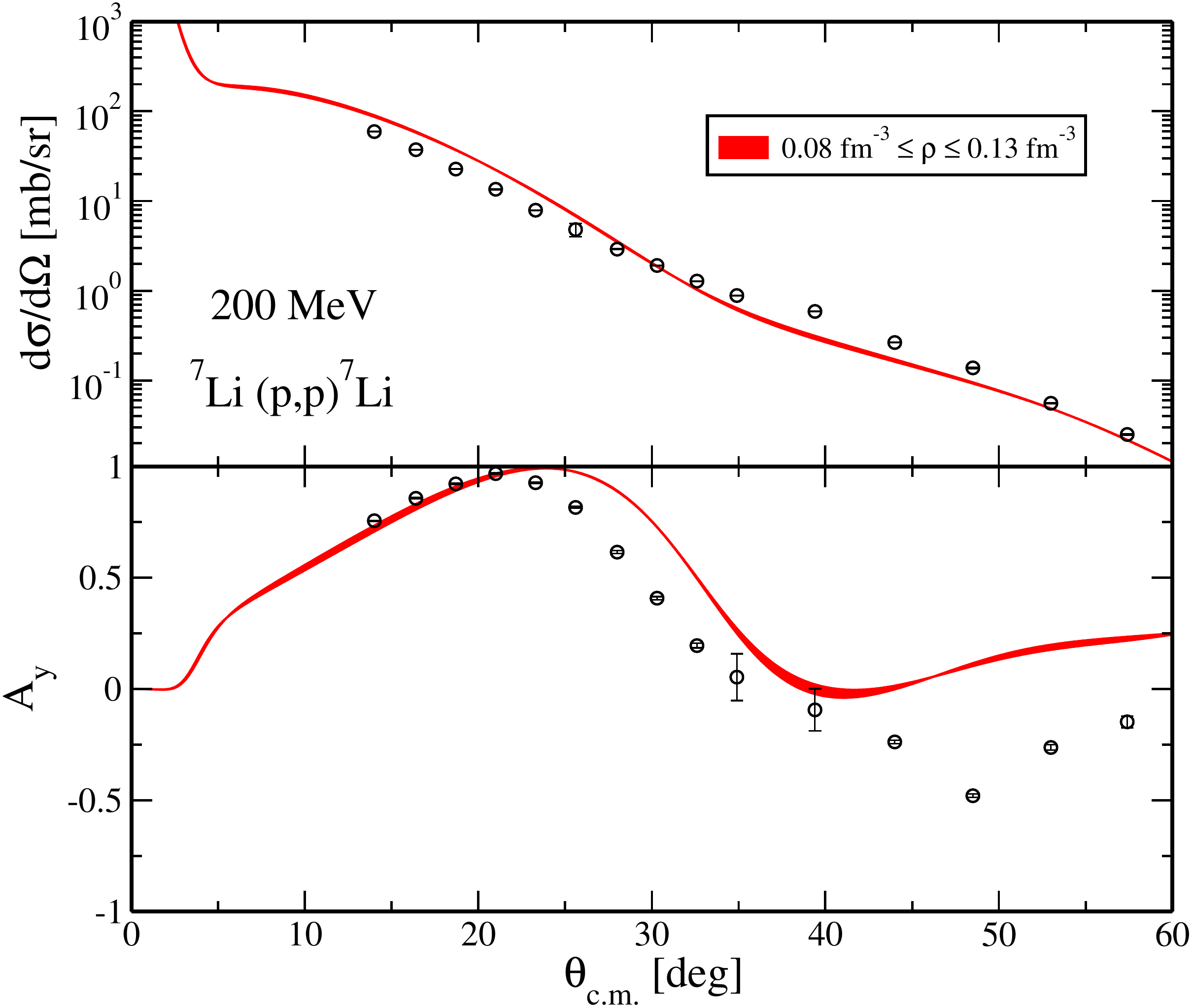}
\end{center}
\caption{\label{fig3}
Differential cross section (upper panel) and analyzing power (lower panel), as 
functions of the center-of-mass scattering angle, for 200 MeV protons elastically 
scattered from $^{7}$Li (J$^{\pi}$ = 3/2$^{-}$).
The results were obtained using the $NN$ $t$ matrix computed with the $NN$ chiral interaction at N\textsuperscript{3}LO order
supplemented by a density dependent NN interaction (where the baryon density is varied in the range between 0.08
 fm$^{-3}$ and 0.13 fm$^{-3}$) and the one-body nonlocal density matrices computed with the NCSM method using
realistic nucleon-nucleon and three-nucleon local-nonlocal chiral interactions.}
\end{figure}

To show the robustness of our proposed approach we will show a couple of non trivial calculations perfomed in the last years. 
The first case is the elastic proton scattering on  $^{7}$Li, in which the target nucleus has spin and parity quantum numbers 
$J^\pi = 3/2^-$. Theoretical predictions along with the experimental data are presented in Fig.~\ref{fig3}.


The differential cross section and analyzing power were measured at the Indiana University Cyclotron Facility using a polarized proton beam at a laboratory bombarding energy of 200.4 MeV \cite{Glover:1991zz}.
The agreement between the theoretical prediction and the empirical data is good for the differential cross section, over all the angular distribution shown in the figure, and satisfactory for the analyzing power for values of the scattering angle up to $\sim 45^{\rm o}$.
 
The second test case is an extension much needed for future applications: inverse kinematics. In fact, in recent years, experimental efforts have multiplied to develop the technologies necessary to study the elastic scattering of protons (and ions) by using inverse kinematics. 
This configuration is necessary to study exotic nuclei that have very short average lifetimes. A very interesting experiment performed at the GSI was the study of the scattering of $^{56,58}$Ni nuclei on hydrogen targets at energies near 400 MeV/nucleon in inverse kinematics for the determination of the distribution of nuclear matter \cite{gsi}. Since the only theoretical approach used so far to analyze the experimental data is the Glauber model which contains phenomenological inputs, it is mandatory to establish a framework in which microscopic calculations could help understading experimental results.

\begin{figure}
\begin{center}
\includegraphics[scale=0.43]{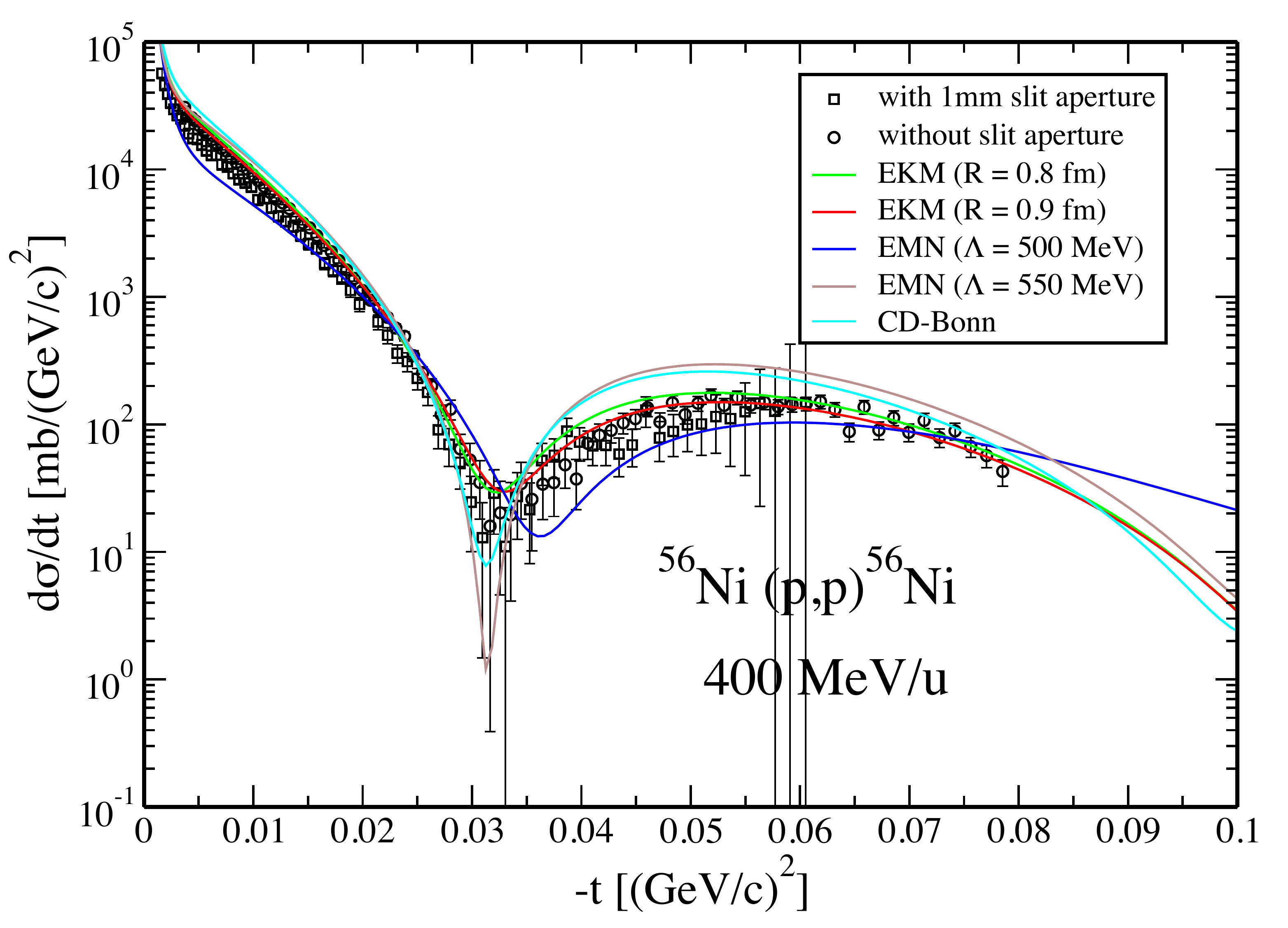}
\end{center}
\caption{\label{fig2} Differential cross section as a function of the Mandelstam variable $t$ in the laboratory frame for $^{56}$Ni(p, p)$^{56}$Ni elastic scattering at $E = 400$ MeV/u. Calculations are performed with different microscopic OPs derived from the $NN$ chiral potentials of Epelbaum 
{\it et al.} (EKM) \cite{Epelbaum:2014sza,Epelbaum:2014efa} and Entem {\it et al.} (EMN) \cite{Entem:2014msa,Entem:2017gor}  at N\textsuperscript{4}LO in comparison with the well known CD-Bonn \cite{Machleidt:2000ge}. With empty black squares and circles are shown the empirical data along with the corresponding error bars \cite{gsi}.}
\end{figure}

In Fig. \ref{fig2} we show the results only for $t \le 0.1$ (GeV/c)$^2$ in the laboratory system because, due to the challenges of inverse kinematics, the regime of the elastic scattering needs to be focused at small center-of-mass angles. The agreement with experimental data is impressive despite the different kinematics and the high-energy regime at which the measurements are performed.


\section{Necessary extensions}

Among the different extensions under development, we would like to mention probably the most relevant one: the inclusion in our framework of
inelastic reactions. 

In the previous sections we mentioned a class of experiments that usually need the subtraction of contributions from the inelastic channel to perform a correct data analysis. In this perspective, if we wish to establish a consistent microscopic approach for inelastic NA scattering, which is our long-term goal, it is mandatory to develop a two-step theory program: first a reliable description of OPs for states with spin-parity quantum numbers J$^\pi \ne 0^+$
(already under study with good results \cite{VorabbiSpin}) and then an extension of our formalism that, so far, is restricted to the elastic channel.

\begin{figure}
\begin{center}
\includegraphics[scale=0.43]{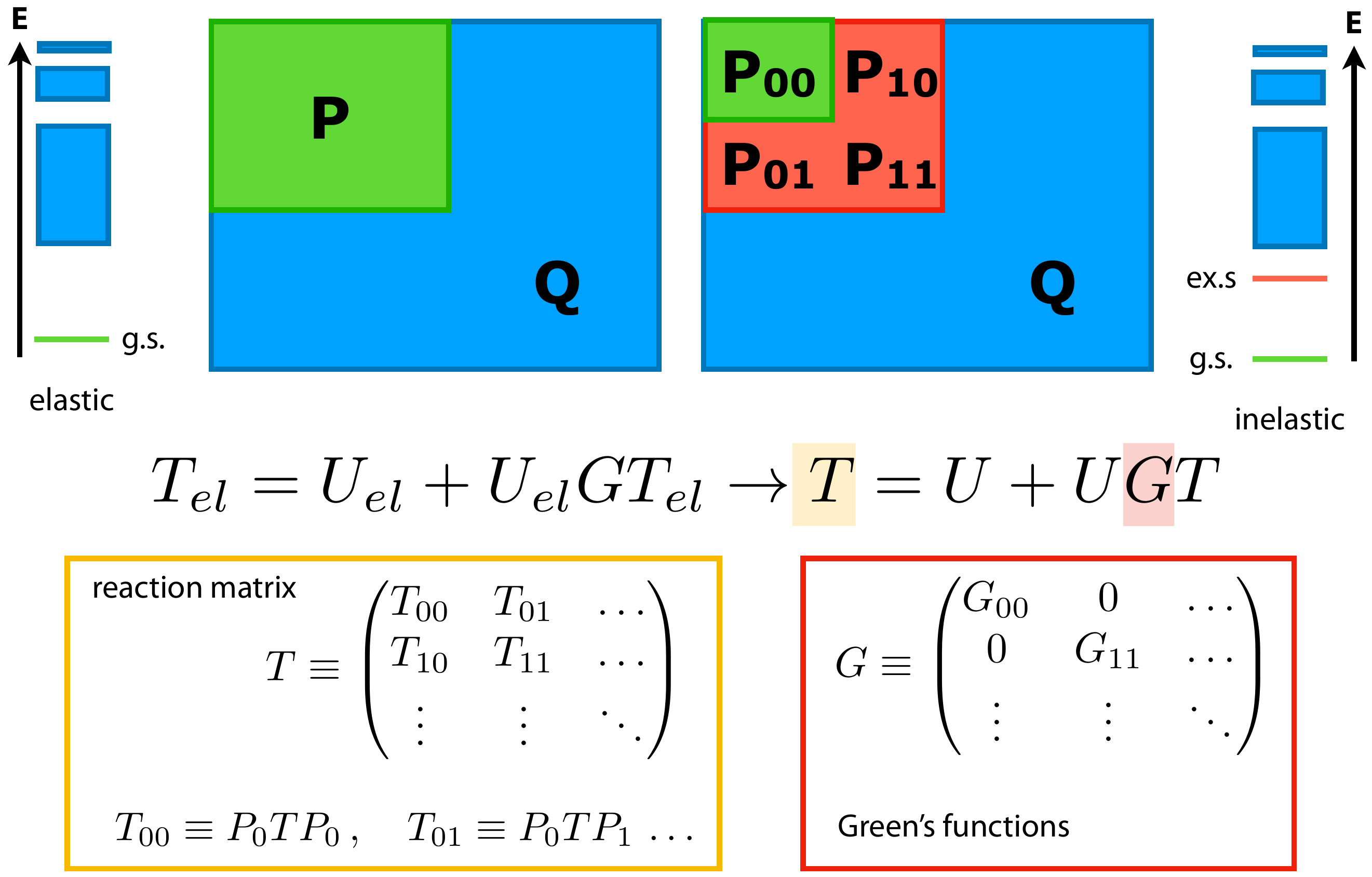}
\end{center}
\caption{\label{fig4} This {\it infographic} shows how the structure of Eq. \ref{generalscatteq} must change in order to include the description of inelastic processes. In order to accommodate the inclusion of the relevant excited state, it is necessary to calculate four optical potentials that describe all the available scattering channels. According to this picture we would need both stationary and transition densities. 
}
\end{figure}

Let's suppose to extend the definition of the projection operator $P$
\begin{equation}
P \equiv P_0 + P_1 = \ket{\Psi_0} \bra{\Psi_0} + \ket{\Psi_1} \bra{\Psi_1} \, ,
\end{equation}
with still $P + Q = \dblone$.
Inserting this relation into the many-body Lippmann-Schwinger equation we could follow the same prescriptions adopted for elastic case to 
define the following quantities
\begin{equation}
T_{00} \equiv P_0 T P_0 \, , \quad T_{01} \equiv P_0 T P_1 \, , \dots \quad G_{00} \equiv P_0 G_0 (E) P_0 \, , \ldots
\end{equation}
that represent the new scattering amplitudes that can be accomodated in a matrix form as follows (see Fig. \ref{fig4})
\begin{equation}
{\bm T} = {\bm U}  + {\bm U}  {\bm G}  {\bm T} \, ,
\end{equation}
with the previous terms defined as
\begin{equation}
{\bm T} \equiv
\begin{pmatrix}
T_{00} & T_{01} \\
T_{10} & T_{11} 
\end{pmatrix}
\, , \qquad
{\bm U} \equiv
\begin{pmatrix}
U_{00} & U_{01} \\
U_{10} & U_{11} 
\end{pmatrix}
\, , \qquad
{\bm G} \equiv
\begin{pmatrix}
G_{00} & 0 \\
0 & G_{11} 
\end{pmatrix}
\, .
\end{equation}
When projected on the momentum basis, the equation for the $00$ component of the scattering matrix becomes
\begin{equation}
T_{00} ({\bm k}^{\prime} , {\bm k}) = U_{00} ({\bm k}^{\prime} , {\bm k})  + \int d {\bm p} \frac{U_{00} ({\bm k}^{\prime} , {\bm p}) T_{00} ({\bm p}, {\bm k})}{E - E(p) + i \epsilon}
+  \int d {\bm p} \frac{U_{01} ({\bm k}^{\prime} , {\bm p}) T_{10} ({\bm p} , {\bm k})}{E - E(p) - e_1 + i \epsilon} \, ,
\end{equation}
and similarly for the other equations. Here $e_1$ is the excitation energy of the first excited state. Of course the previous equation has to be expanded in partial waves along with the other 3 equations. Work along this direction is in progress \cite{VorabbiNext}.


\end{document}